\documentclass{iopart}

\usepackage{iopams}  
\usepackage{graphicx}
\usepackage{subfig}
\usepackage{subfloat}
\usepackage{color}

\begin{document}

\title[]{Zooming into market states}

\author{Desislava Chetalova, Rudi Sch\"afer and Thomas Guhr}

\address{Fakult\"at f\"ur Physik, Universit\"at Duisburg--Essen, D--47048 Duisburg, Germany}
\ead{desislava.chetalova@uni-due.de}

\begin{abstract}
We analyze the daily stock data of the Nasdaq Composite index in the 22-year period $1992-2013$ and identify market states as clusters of correlation matrices with similar correlation structures. 
We investigate the stability of the correlation structure of each state by estimating the statistical fluctuations of correlations due to their non-stationarity. Our study is based on a random matrix approach recently introduced to model the non-stationarity of correlations by an ensemble of random matrices. 
This approach reduces the complexity of the correlated market to a single parameter which characterizes the fluctuations of the correlations and can be determined directly from the empirical return distributions. 
This parameter provides an insight into the stability of the correlation structure of each market state as well as into the correlation structure dynamics in the whole observation period.
The analysis reveals an intriguing relationship between average correlation and correlation fluctuations. The strongest fluctuations occur during periods of high average correlation which is the case particularly in times of crisis. 
\end{abstract}

\maketitle

\section{Introduction}

Financial markets are highly complex and continuously evolving systems. To understand their statistical behavior and dynamics the analysis of the correlations between the market constituents is of crucial importance. 
Much research has thus been focused on the information acquired by the correlations, see e.g., \cite{Stanley2000, Potters2000}.

Recently, correlation matrices were used to identify states of a financial market based on similarities in the correlation structure at different times \cite{Muennix2012}. The study revealed the existence of several typical market states between which
the market jumps back and forth over time. 
Each market state has a characteristic correlation structure and temporal behavior. 

Here, we take a closer look at the statistics of market states. In particular, we investigate the stability of their corresponding correlation structures. We address this question by means of a random matrix approach introduced in \cite{Schmitt2013, Chetalova2013}. 
It models the non-stationarity of the correlations by an ensemble of random matrices.
This approach not only yields a quantitative description of heavy tailed return distributions but also reduces the effect of fluctuating correlations to a single parameter measuring the fluctuation strength. 
Our random matrix approach provides a method to estimate the fluctuations of the actual correlations due to non-stationarity \cite{Bekaert1995, Longin1995, Fenn2011}. This is a crucial difference to another usage of random matrices where the estimation errors which arise due to the finiteness of the time series are modeled \cite{Laloux1999, Plerou1999, Potters2005, Burda2006, Abul2009, Burda2011}.

We note that our study is based on the definition of market states as clusters of correlation matrices, as first suggested in \cite{Muennix2012}. The concept of different market states or regimes in which the market operates is, however,
not entirely new to the economics literature, see e.g., \cite{Schaller1997, Marsili2002}.

The paper is organized as follows. In section~\ref{section2}, we identify market states for the Nasdaq Composite stock market in the period $1992-2013$ by performing a clustering analysis based on the Partitioning Around Medoids (PAM) algorithm \cite{Kaufman1990} and study their dynamics. 
We briefly summarize the main features of the random matrix model in section~\ref{section3}. In section~\ref{section4}, we apply its results to study the stability of the correlation structure of each market state as well as the correlation structure dynamics in the whole 
observation period. We conclude our findings in section~\ref{section5}.

\section{Market states: Identification and dynamics}   \label{section2}

We begin with identifying the market states as clusters of correlation matrices. We consider $K=258$ stocks of the Nasdaq Composite index traded in the 22-year period from January 1992 to December 2013 \cite{Yahoo} 
which corresponds to  5542 trading days. 
For each stock $k$ we calculate the return time series
\begin{equation}
 r_k (t) = \frac{ S_k(t + \Delta t)-S_k(t) }{ S_k(t) } \ , \quad k=1,\dots, K \ ,
\label{returns}
\end{equation}
where $ S_k(t) $ is the price of the $ k $-th stock at time $ t $ and $ \Delta t $ is the return interval. We choose $\Delta t $ to be 1 trading day and calculate the daily returns for each stock.

The correlations between time series are commonly measured via the Pearson correlation coefficient
\begin{equation}
 C_{kl}  = \frac{\langle r_k(t) r_l(t) \rangle - \langle r_k(t) \rangle \langle r_l(t) \rangle }{ \sqrt{\langle r_k^2(t) \rangle - \langle r_k(t) \rangle ^2} \sqrt{\langle r_l^2(t) \rangle - \langle r_l(t) \rangle ^2} } \ ,
\label{pearson}
\end{equation}
where $\langle \dots \rangle$ denotes the average over a time window yet to be specified. The main object of interest in the following is the $K \times K $ correlation matrix $C$ which contains the correlation coefficients between all pairs of return time series.

The Pearson correlation coefficient is commonly used as a measure of dependence. Sometimes, however, it can be problematic in particular for non-linear dependencies or for non-stationary data.
The latter is extremely relevant for financial data since drift and volatilities \cite{black76, christie82, Schwert1989} fluctuate considerably in time. Thus, the correlation coefficient averages over time-varying trends and volatilities, which results in an
estimation error of the correlations. 
In order to eliminate this kind of error we employ the method of local normalization \cite{Schaefer2010}.
For each return time series we subtract the local mean and divide by the local standard deviation
\begin{equation}
\hat r_k (t) = \frac{ r(t)-\langle r(t)\rangle_n }{ \sqrt{\langle r^2(t) \rangle_n - \langle r(t)\rangle^2_n} } \quad ,
\end{equation}
where $\langle \dots \rangle_n$ denotes the local average which runs over the $n$ most recent sampling points. For daily data we use $n=13$ as discussed in \cite{Schaefer2010}. 
Thus, the local normalization removes the local trends and variable volatilities while preserving the correlations between the time series. An alternative approach would be to use the residuals of a GARCH fit \cite{Bollerslev1986}, which also yields stationary time series. We choose the local 
normalization, as it does not require any model assumptions.

\begin{figure*}[h]
\centering
\includegraphics[width=\textwidth]{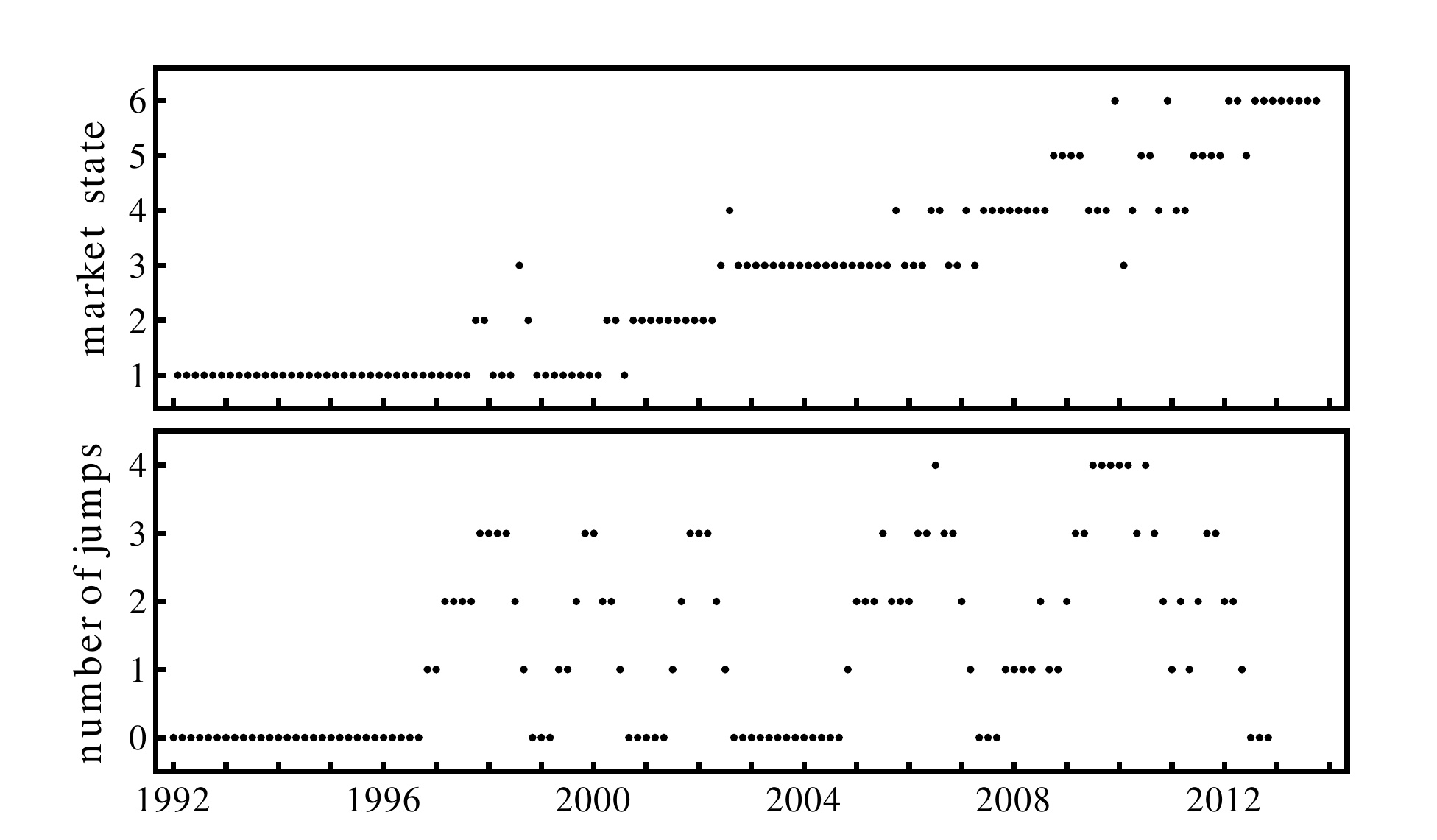}
\caption{Top: Time evolution of the market in the observation period $1992-2013$. Each point represents a correlation matrix measured over a two-month time window. Bottom: Number of jumps between states calculated on a one-year sliding window. 
The first point represents the number of jumps in the period $1/92-12/92$, the second point--in the period $3/92-2/93$ and so on.}
\label{fig1}
\end{figure*}

Using the locally normalized daily returns we now obtain a set of correlation matrices measured on disjoint two-month intervals of the 22-year observation period. To identify the market states we perform a clustering analysis based on the PAM algorithm where the number of clusters is estimated via the gap statistic \cite{Tibs2001}.
The clustering analysis separates the correlation matrices into 6 clusters based on the similarity of their correlation structures. Each cluster is associated with a market state.
Figure~\ref{fig1} (top) shows the time evolution of the market states. 
The market switches back and forth between states: Sometimes it remains in a state for a long time, sometimes it jumps briefly to another state and returns back or evolves further. 
On longer time scales, the market evolves towards new states, whereas previous states die out.
How frequent does the market switch between states? Figure~\ref{fig1} (bottom) shows the number of jumps from one state to another calculated on a one-year sliding window. 
\begin{figure*}[pt]
\centering
\subfloat[]{\includegraphics[width=0.48\textwidth]{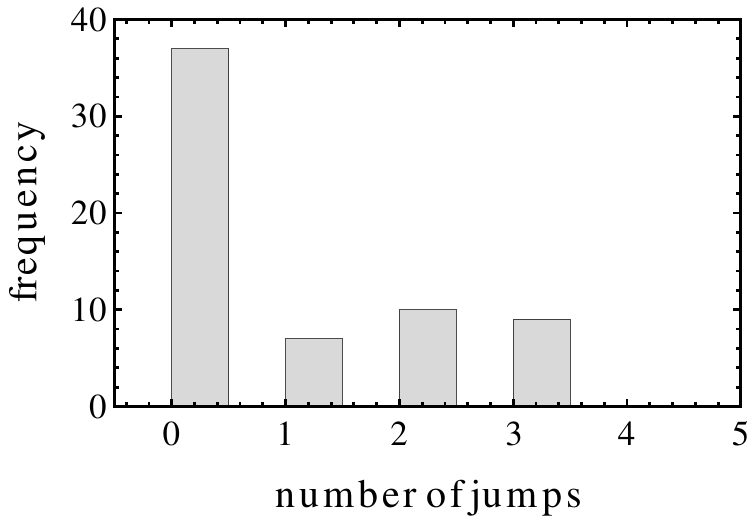}}
\quad
\subfloat[]{\includegraphics[width=0.48\textwidth]{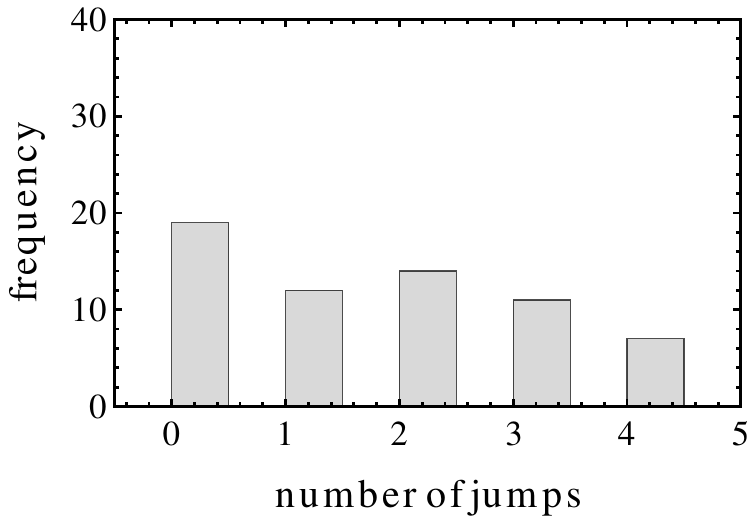}}
\caption{ Histograms of the number of jumps between states (a) in the first half $1992-2002$ and (b) in the second half $2003-2013$ of the observation period.}
\label{fig2}
\end{figure*}
After a stable 5-year period we observe that the market begins to switch between states. The highest number of jumps per year can be found in the period around 2010. In figure~\ref{fig2} we compare the number of jumps frequency in both halves of the observation period.
In the second half of the observation period the number of jumps per year increases. At the same time, the lifetime, i.e., the time the market stays in a certain state before it jumps to another one, decreases.
Figure~\ref{fig3} shows the histograms of the lifetime in both halves of the observation period. The first half contains mostly long-lived states. In the second half the frequency of the short-lived states increases considerably, while the frequency of the long-lived states decreases.
\begin{figure*}[pt]
\centering
\subfloat[]{\includegraphics[width=0.48\textwidth]{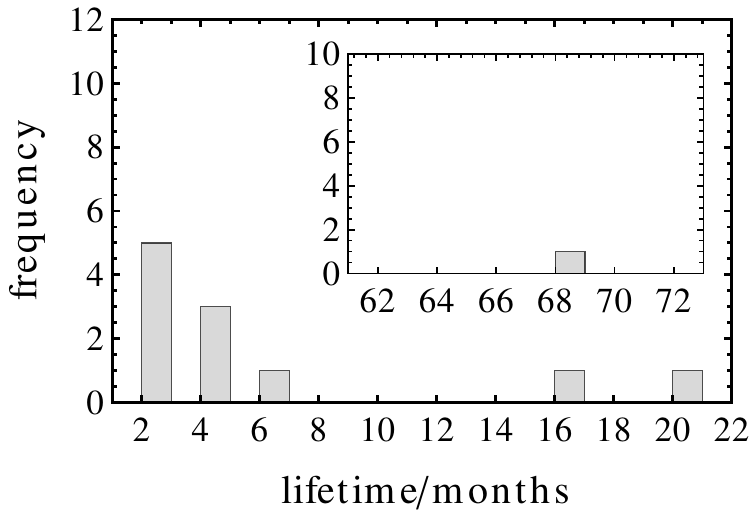}}
\quad
\subfloat[]{\includegraphics[width=0.48\textwidth]{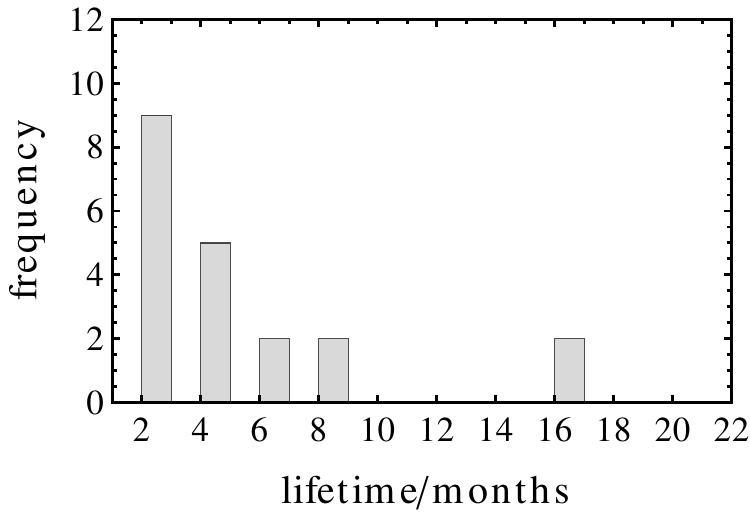}}
\caption{ Histograms of the lifetime in months (a) in the first half $1992-2002$ and (b) in the second half $2003-2012$ of the observation period.}
\label{fig3}
\end{figure*}

\begin{figure*}[p]
\subfloat[state 1]{\includegraphics[width=0.35\textwidth]{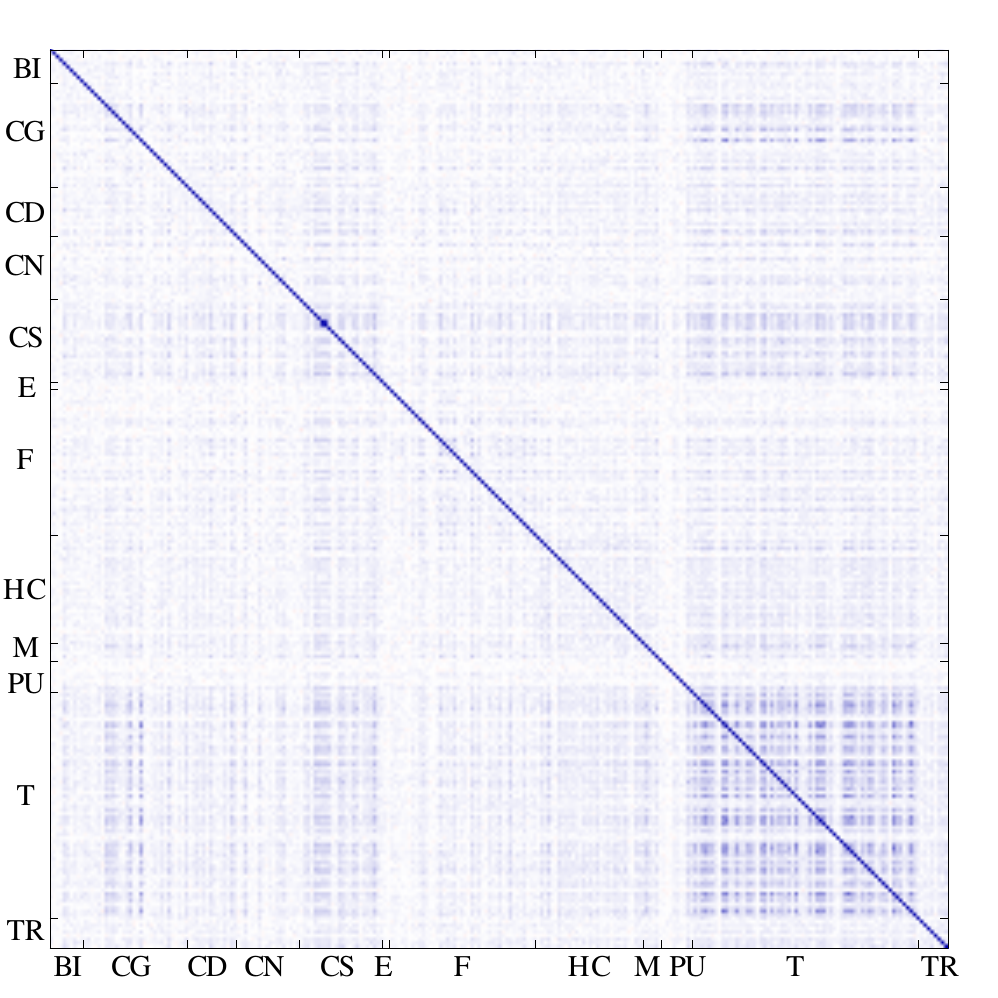}}
\subfloat[state 2]{\includegraphics[width=0.35\textwidth]{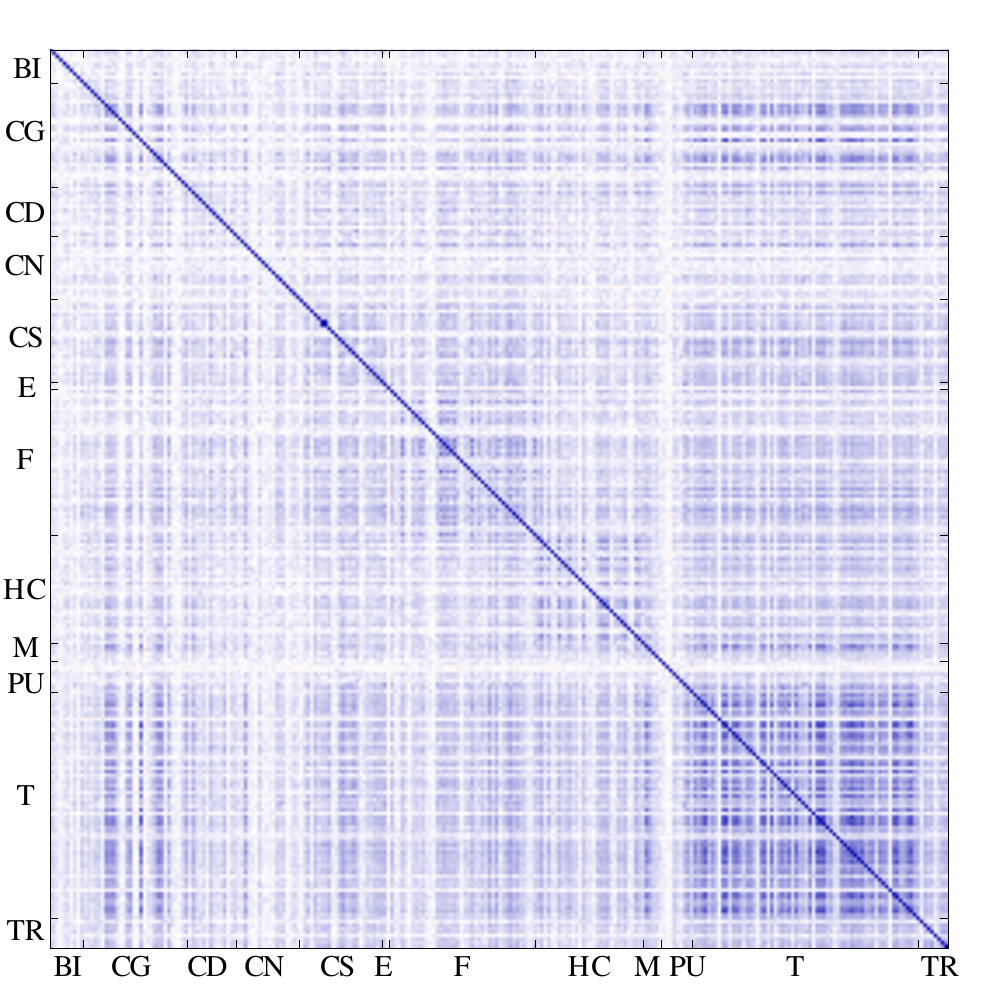}}
\subfloat[state 3]{\includegraphics[width=0.35\textwidth]{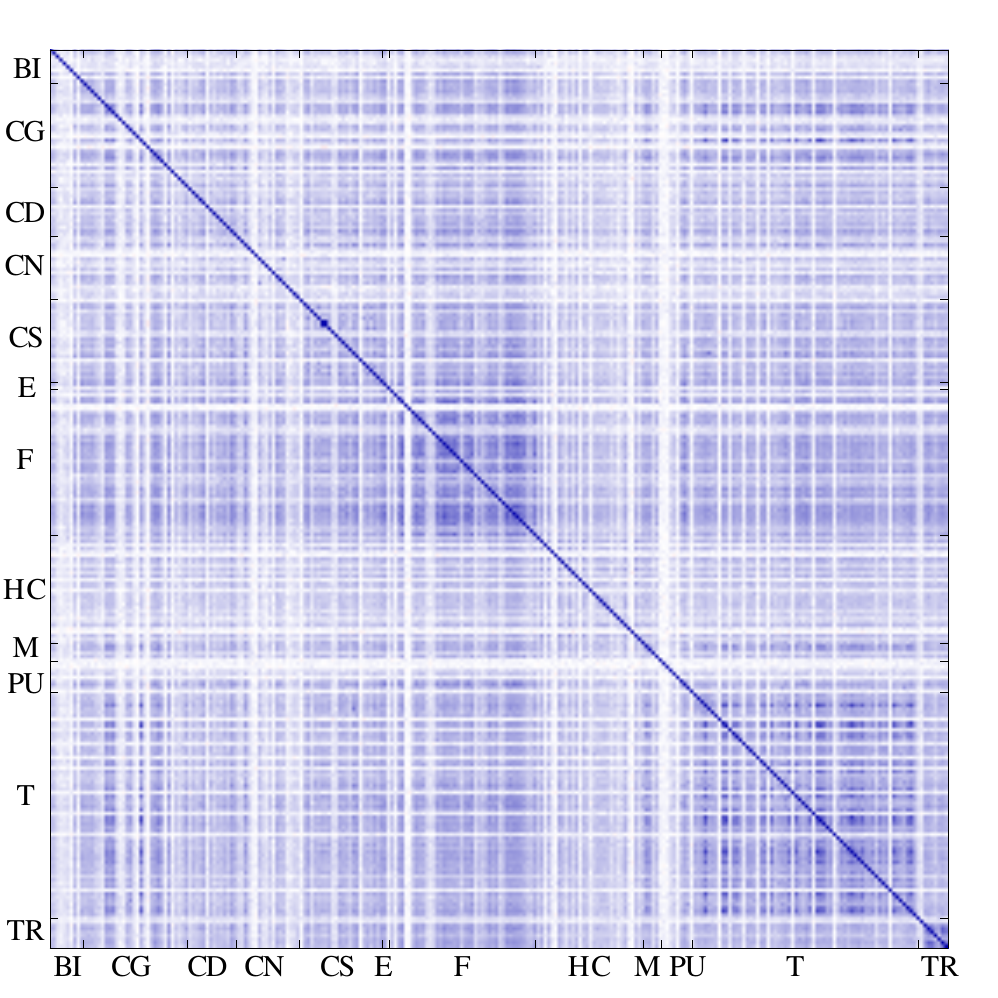}}
\\
\subfloat[state 4]{\includegraphics[width=0.35\textwidth]{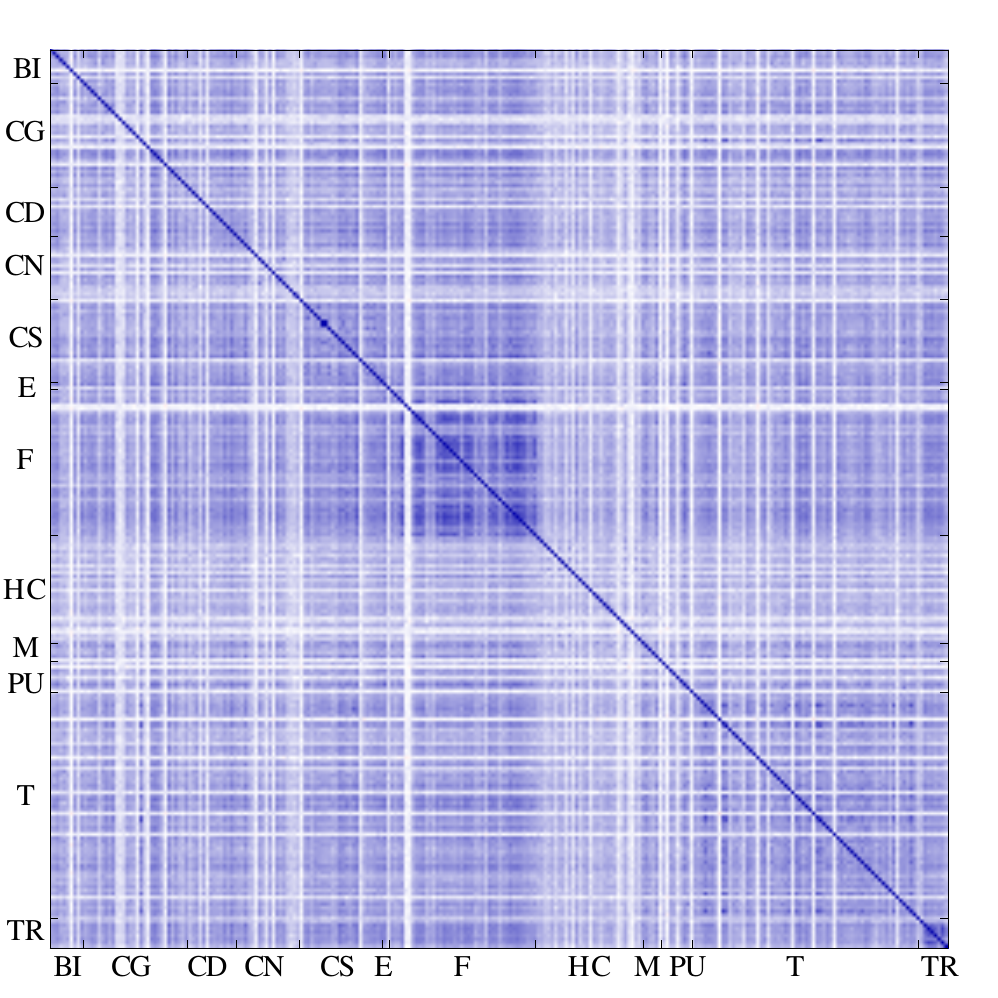}}
\subfloat[state 5]{\includegraphics[width=0.35\textwidth]{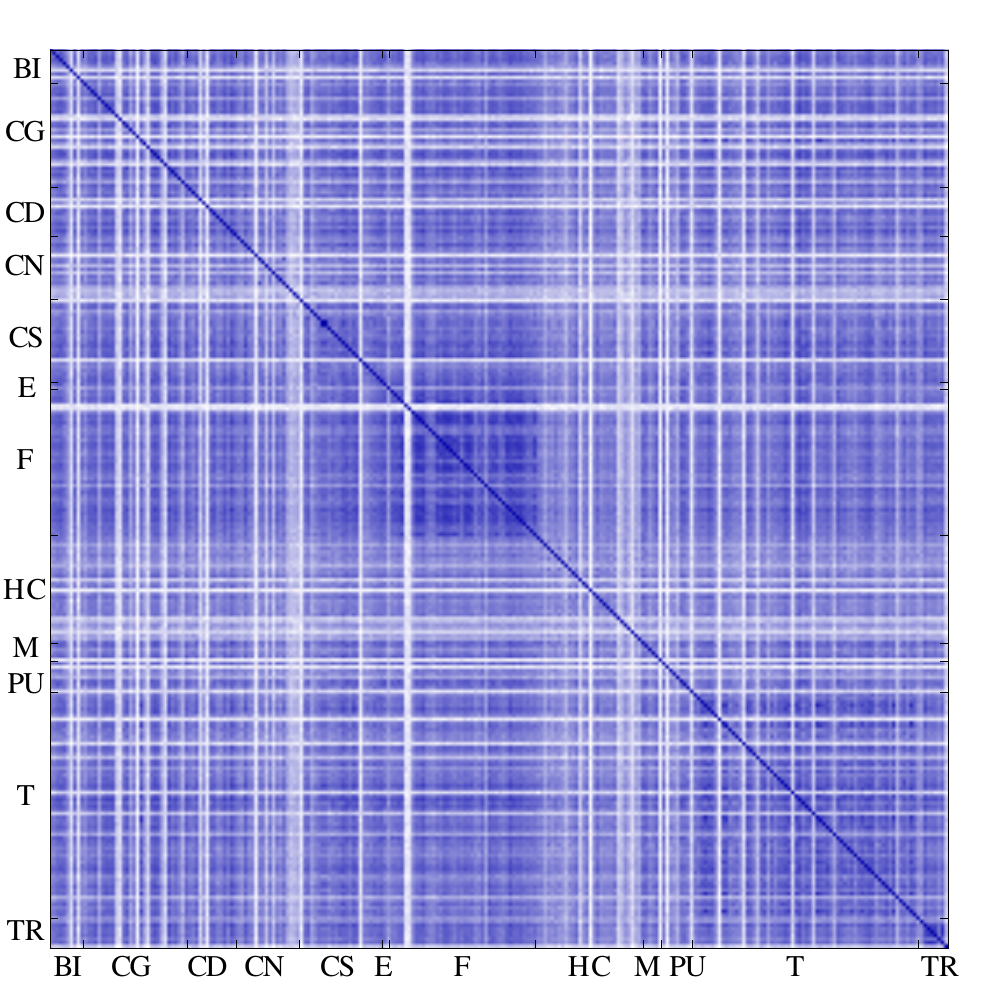}}
\subfloat[state 6]{\includegraphics[width=0.35\textwidth]{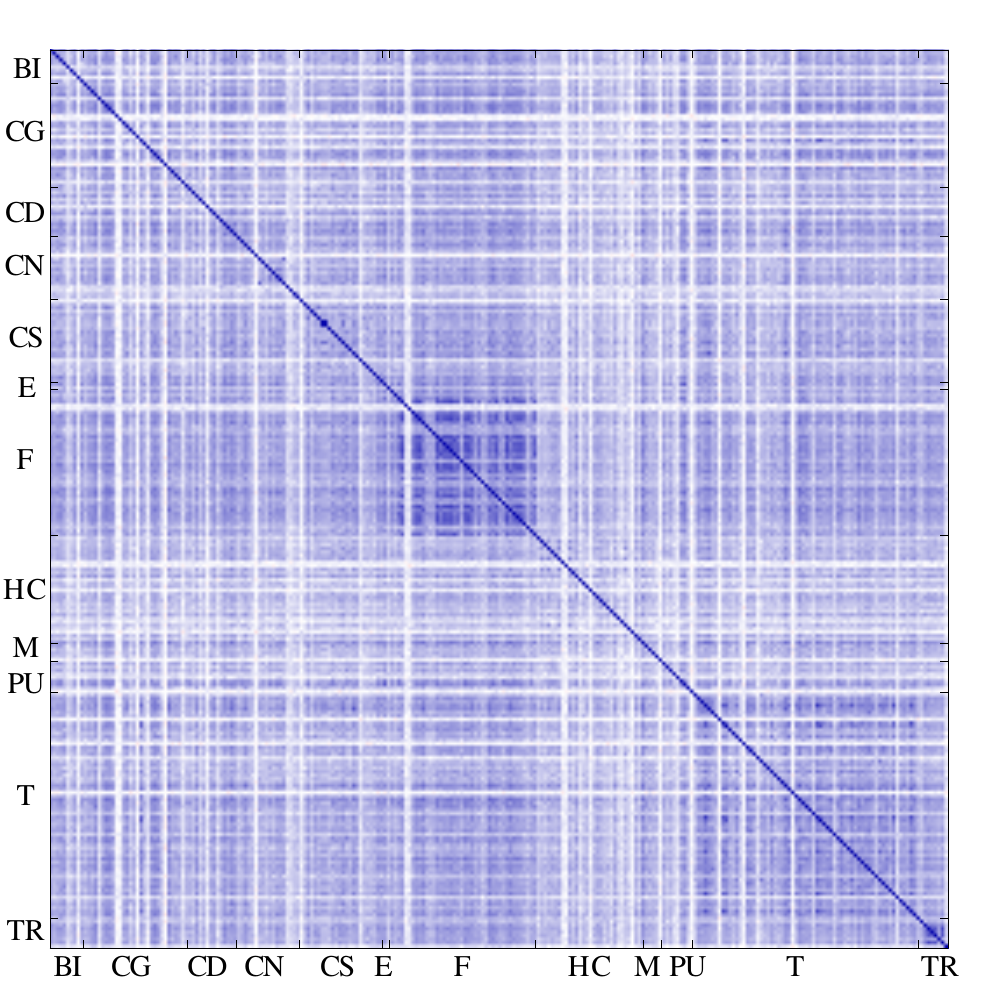}}
\\
\centering \subfloat[Average
  correlation matrix]{\includegraphics[width=0.65\textwidth]{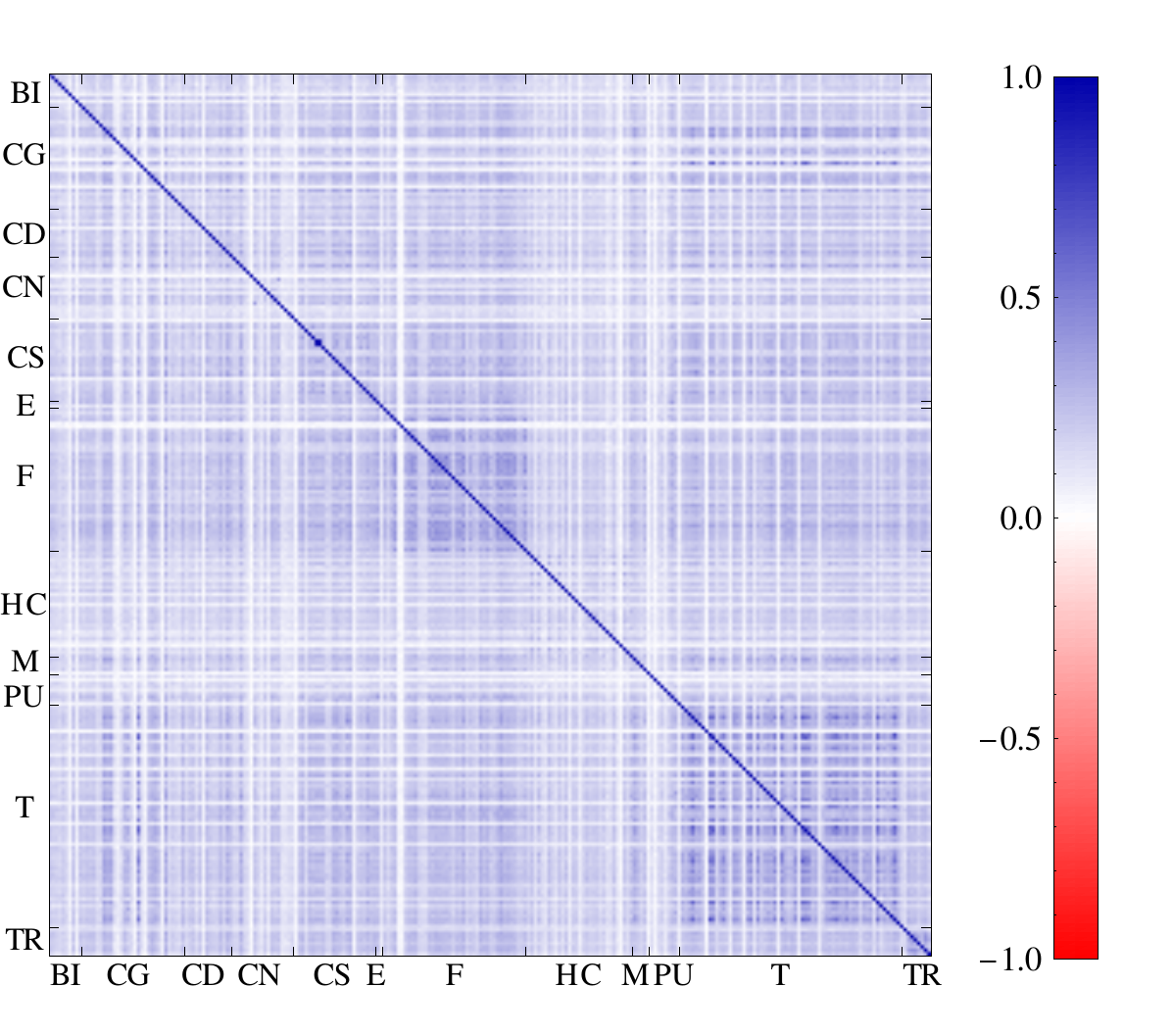}}     
\caption{(a)--(f) Average correlation matrices for each market state. (g) Overall average correlation matrix. Industry sectors legend: BI: Basic Industries, CG: Capital Goods, 
CD: Consumer Durables, CN: Consumer Non-Durables, CS: Consumer Services, E: Energy, F: Finance, HC: Health Care, M: Miscellaneous, PU: Public Utilities, 
T: Technology, TR: Transportation.}
\label{fig4}
\end{figure*}

To illustrate the different correlation structures of each state, we sort the stocks according to their industry sector and calculate the corresponding average correlation matrices, see figure~\ref{fig4}. We indeed recognize different characteristic
correlation structures. State 1 shows an overall weak correlation. In state 2 we have the strongest correlation within the technology sector and between technology and capital goods, whereas in states 3 and 4 the correlation within the finance sector is the strongest.
We observe that the average correlation level increases from state to state, reaching its highest value in state 5. In state 6 the average correlation level decreases. The finance sector, however, is still strongly correlated. 
Further, we note that the health care sector is weakly correlated to the rest of the market in almost all states. 

\section{Random matrix approach to non-stationary correlations}   \label{section3}

Before we take a closer look at each market state we briefly summarize the main aspects of our random matrix approach, which gives us a handy tool to estimate the correlation fluctuations due to non-stationarity.

\begin{figure*}[ht]
\centering
\subfloat[]{\includegraphics[width=0.48\textwidth]{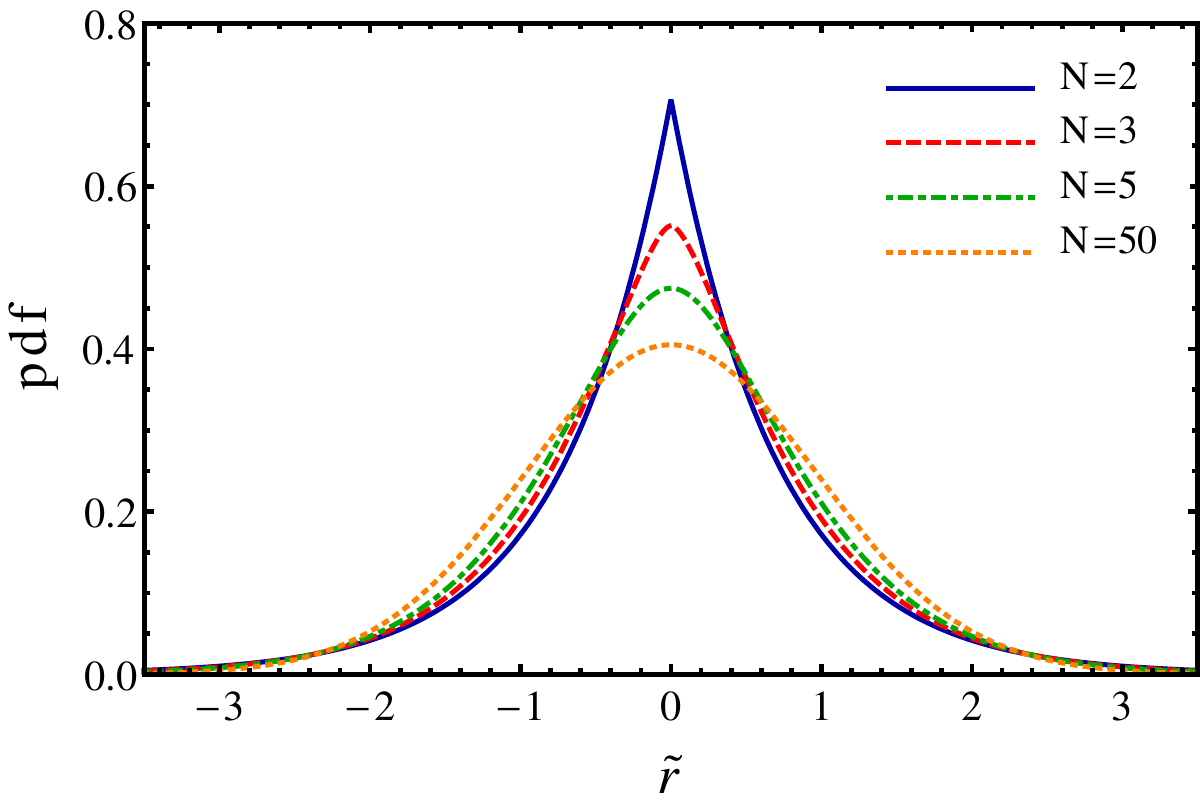}}
\hspace*{2mm}
\subfloat[]{\includegraphics[width=0.49\textwidth]{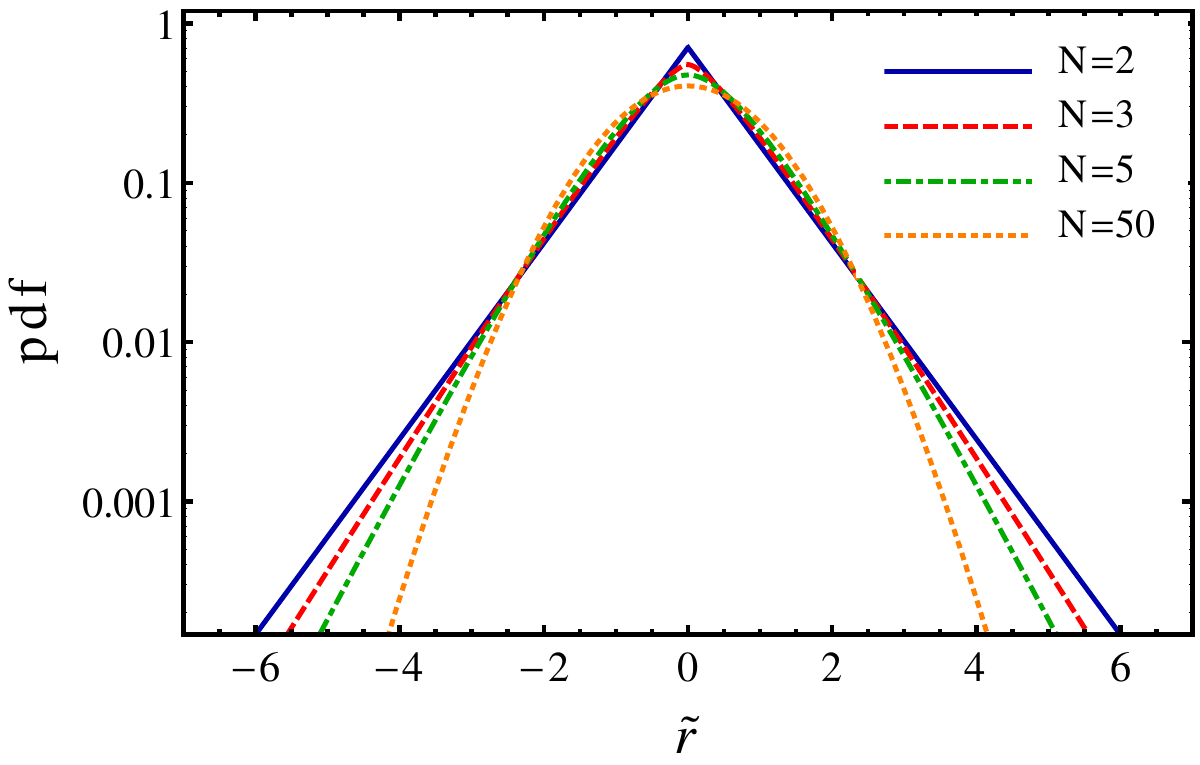}}
\caption{Illustration of the average return pdf $\langle g \rangle (\tilde r| N)$ for different values of $N$, plotted  (a) linearly and (b) logarithmically. Solid, dashed, dashed-dotted and 
dotted lines correspond to $N=2, 3, 5, 50$, respectively.}
\label{fig5}
\end{figure*}

In Ref.~\cite{Chetalova2013} we introduce a random matrix approach to model time-varying correlations in financial time series. For the convenience of the reader we sketch the salient features.
We consider a correlated market consisting of $K$ stocks. At each time $t$ we may assume the $K$ component return vector $ r (t) = \left( r_1 (t),\dots ,r_K (t) \right)$ to be normally distributed with a correlation matrix $C_t$.
We checked
and verified this assumption \cite{Schmitt2013}.
 We take the non-stationarity of the correlations into account by replacing the correlation matrix at each time $t$ by a random matrix
\begin{equation}
 C_t \quad \longrightarrow \quad  W W^\dagger  \ ,
\end{equation}
where the dagger denotes the transpose.
The elements of the $K\times N$ rectangular random matrix $ W $ are drawn from a Gaussian distribution with the probability density function (pdf)
\begin{equation}
w ( W | C ,N) = \sqrt { \frac{ N }{ 2 \pi } }^{ K N } \frac{ 1 }{ \sqrt{\det C }^N } \exp \left( - \frac{ N }{ 2 } \tr  W^\dagger  C^{-1}  W \right) \ , 
\label{wishart}
\end{equation}
where $C$ is the empirical average correlation matrix, computed over the whole observation period, not to be confused with the matrices $C_t$ introduced above.
Hence, we model the non-stationary correlation matrices by an ensemble of random Wishart matrices $ WW^\dagger$ which fluctuate around the empirical correlation matrix $C$. 
The variance of the Wishart ensemble is determined by the empirical correlation matrix scaled with the parameter $N$
\begin{equation}
 {\rm var} \left( \left[ WW^\dagger \right]_{kl} \right) =\frac{1 + C^2_{kl}}{N} \ ,
\end{equation}
where $C_{kl}$ is the $kl$-th element of $C$. Thus, the parameter $N$ is directly related to the fluctuation strength of the correlations. The larger $N$, the smaller the fluctuations around $C$ become, eventually vanishing in the limit
$N \rightarrow \infty$.
Averaging over the random matrix ensemble we derive a correlation averaged multivariate normal distribution for the returns of a correlated financial market. The average return pdf reads
\begin{equation}
\langle  g \rangle ({r}| \Sigma, N)  = \frac{ \sqrt{2}^{ 2-N } \sqrt{N}^K }{ \Gamma(N/2) \sqrt{ \det (2 \pi \Sigma) } } \frac{ \mathcal{K}_{\frac{K-N}{2}} \left( \sqrt{ N r^\dagger \Sigma^{-1} r} \right) }{ \sqrt{ N  r^\dagger \Sigma^{-1} r}^{\frac{K-N }{ 2 } } }  \ ,
\label{mresult}
\end{equation}
where $\mathcal K_\nu$ is the modified Bessel function of the second kind of order $\nu$. It depends only on the empirical covariance matrix $\Sigma = \sigma C \sigma$, evaluated over the whole observation period, where $\sigma$ is the diagonal matrix of the volatilities $\sigma_k$.
The free parameter $N$ can be determined by fitting to the data.
Thus, we can assess the correlation fluctuations, which provide information about the stability of the correlation structure in the considered observation period, directly from the empirical data. For the comparison with empirical returns we rotate the return vector $r$ into the eigenbasis of the covariance matrix and normalize its components with the eigenvalues. 
Integrating out all but one component of the rotated vector, which we call $\tilde r$, leads to 
\begin{equation}
\langle g \rangle (\tilde r | N) = \frac{ \sqrt{ 2 }^{ 1-N } \sqrt{N}} { \sqrt{ \pi } \ \Gamma(N/2) } \sqrt{ N \tilde r^2  } ^{ \frac{ N-1 }{ 2 } } \ \mathcal K_{ \frac{ N-1 }{ 2 } }\left( \sqrt{ N \tilde r^2} \right) \ .
\label{result}
\end{equation}
We illustrate this pdf for different values of $N$ in figure~\ref{fig5}. It has exponential tails, which become more and more dominant the smaller $N$. The smaller $N$ the stronger the fluctuations around $C$.
For large $N$ the pdf approaches the normal distribution. The limit corresponds to a stationary case with no fluctuations around $C$.

We note that in the calculation of the average return distribution we treated the volatility $\sigma$ as fixed. However, we can alternatively model the full covariance matrix by an ensemble of random Wishart matrices, which leads to exactly the same result (\ref{mresult}). In this case,
$N$ can also be interpreted as fluctuation strength around the average covariance matrix $\Sigma=\sigma C \sigma $. In the following we use the interpretation of $N$ as fluctuation strength around $C$, which we justify in the next section.

\section{Correlation structure: Stability and dynamics} \label{section4}

The random matrix approach provides a tool to estimate the fluctuation strength of correlations in a given time interval directly from the empirical return distribution.
This allows us to investigate the stability of the correlation structure for a given market state.

\begin{figure*}[p]
\subfloat[state 1: $N=12.8$]{\includegraphics[width=0.36\textwidth]{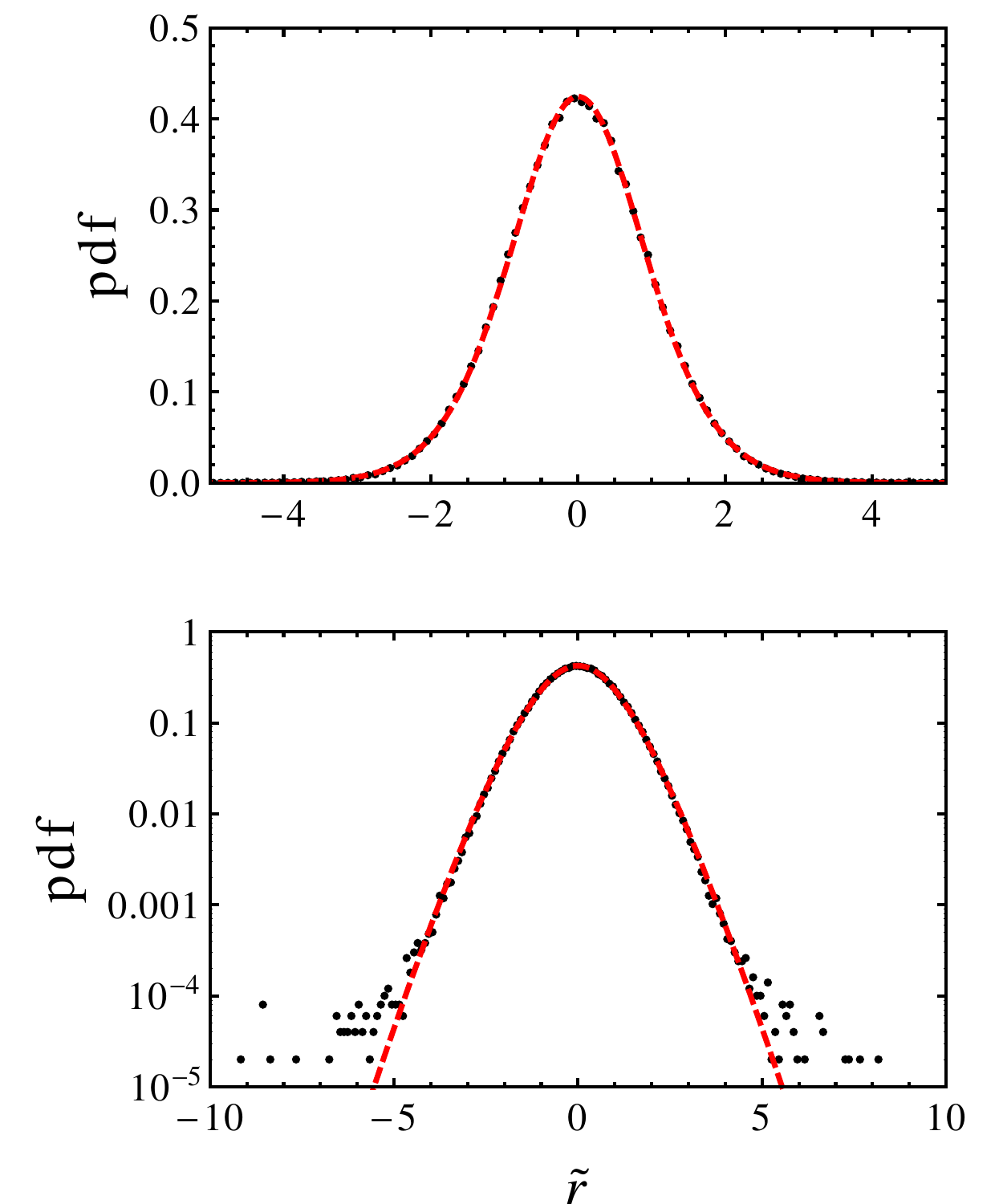}}
\subfloat[state 2: $N=13.4$]{\includegraphics[width=0.36\textwidth]{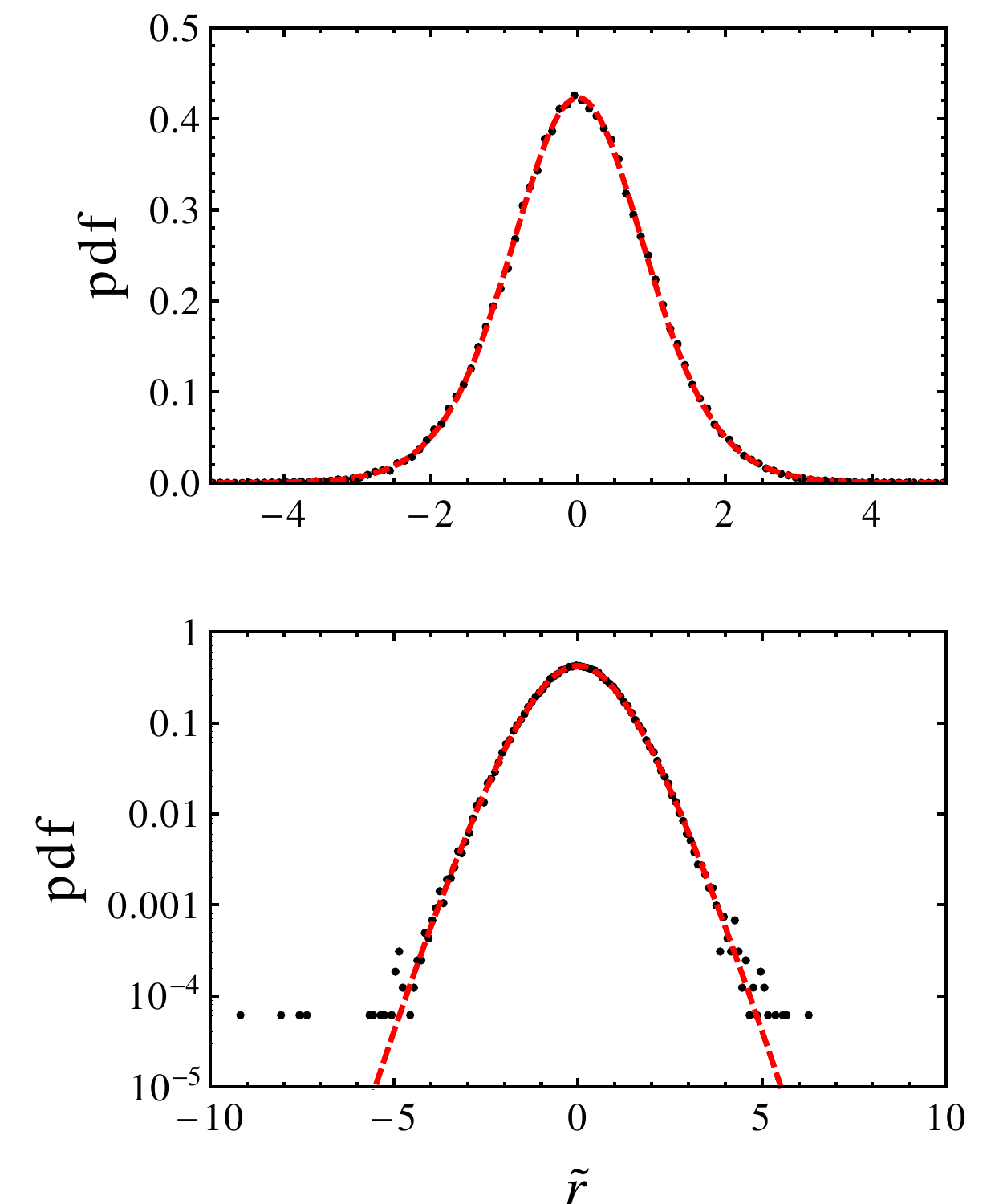}}
\subfloat[state 3: $N=7.8$]{\includegraphics[width=0.36\textwidth]{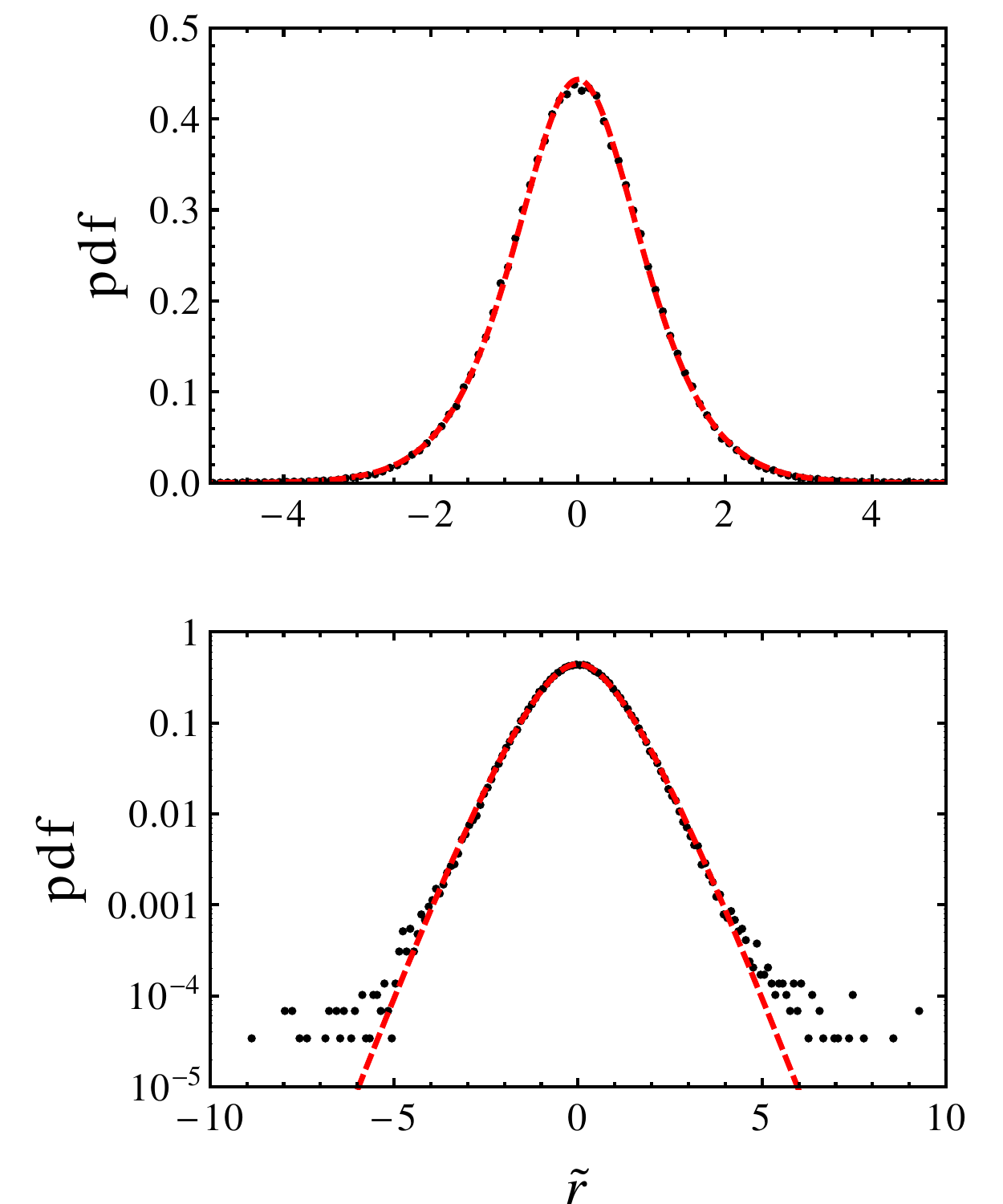}}
\\
\subfloat[state 4: $N=7.8$]{\includegraphics[width=0.36\textwidth]{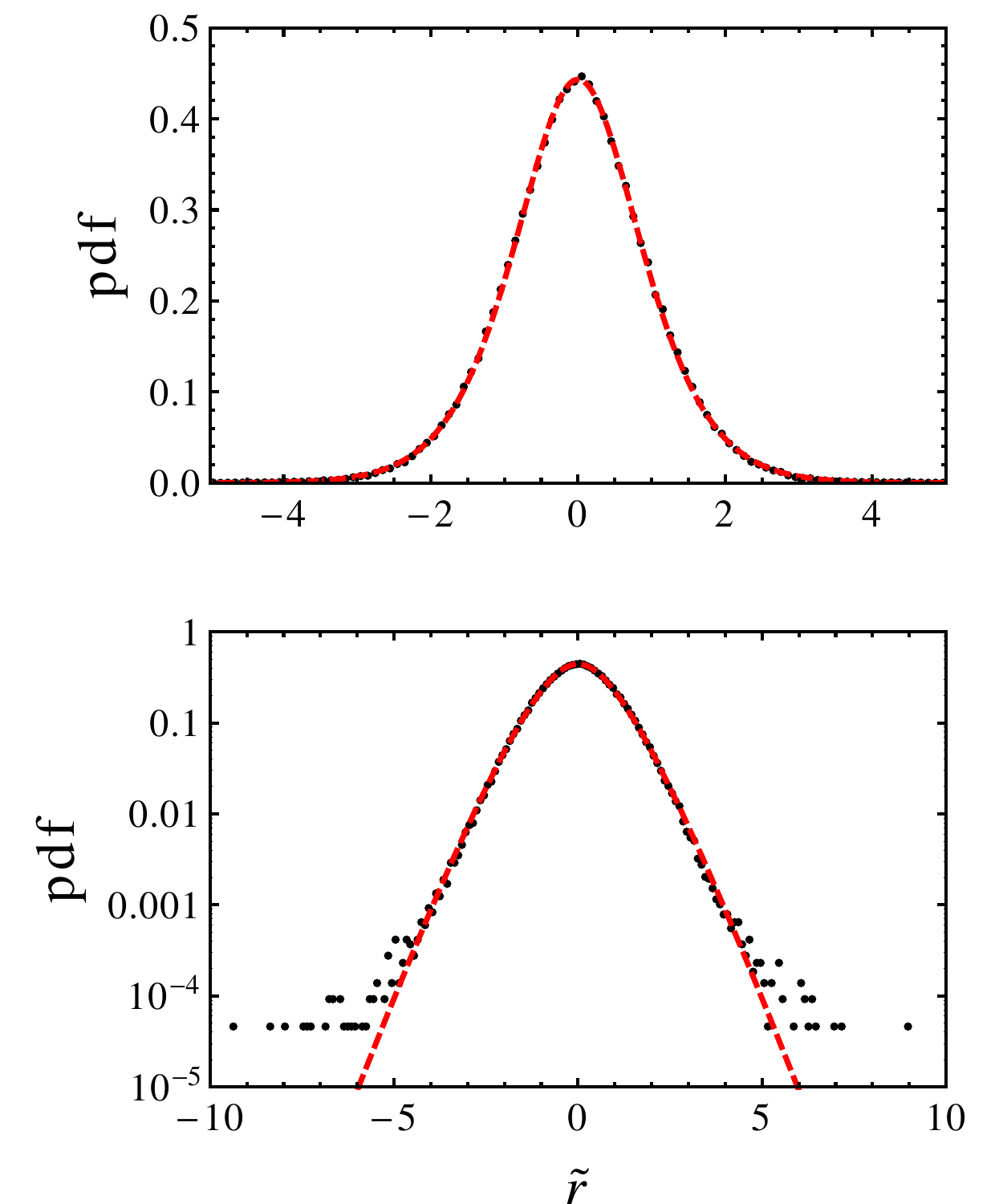}}
\subfloat[state 5: $N=6.1$]{\includegraphics[width=0.36\textwidth]{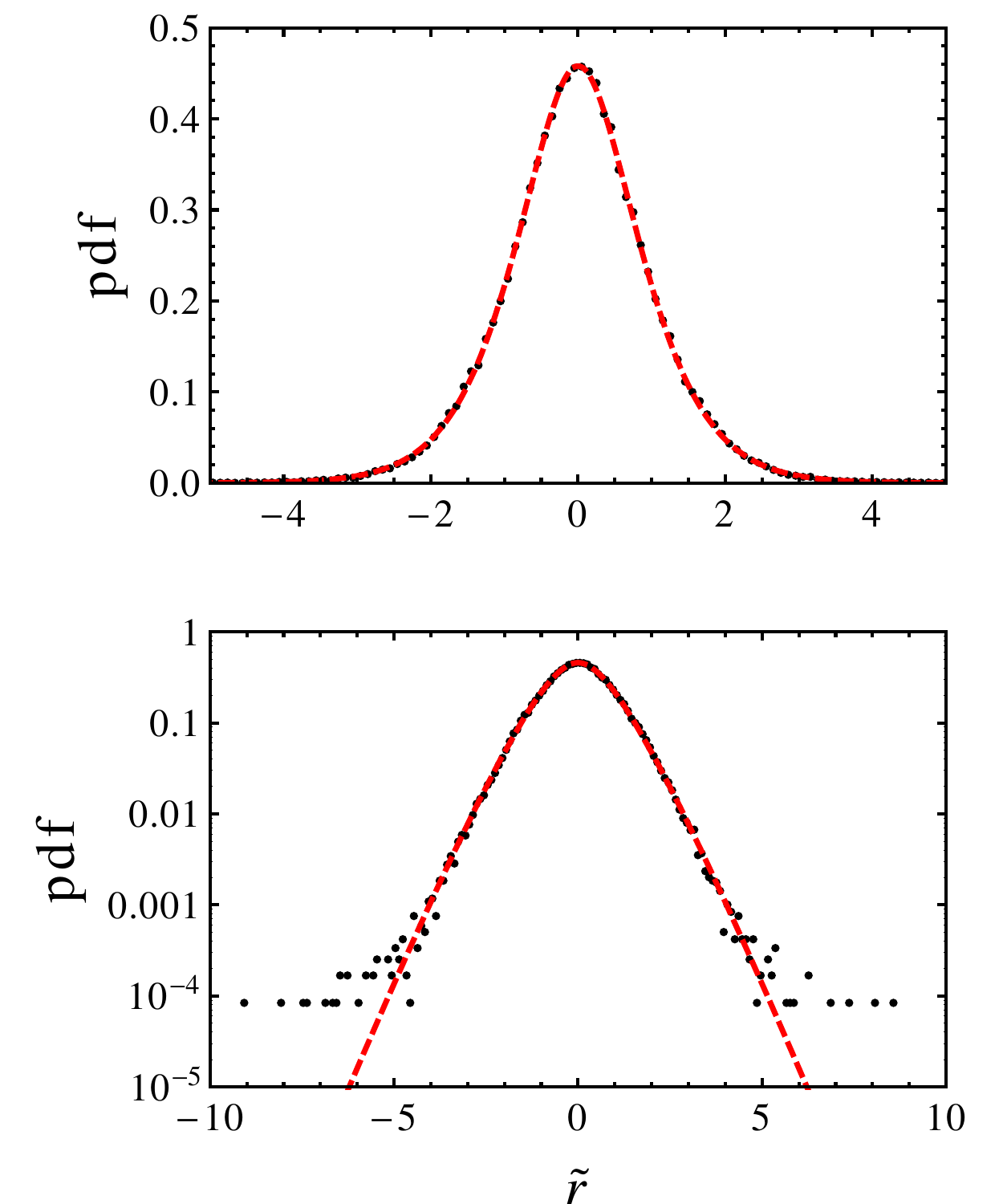}}
\subfloat[state 6: $N=12.0$]{\includegraphics[width=0.36\textwidth]{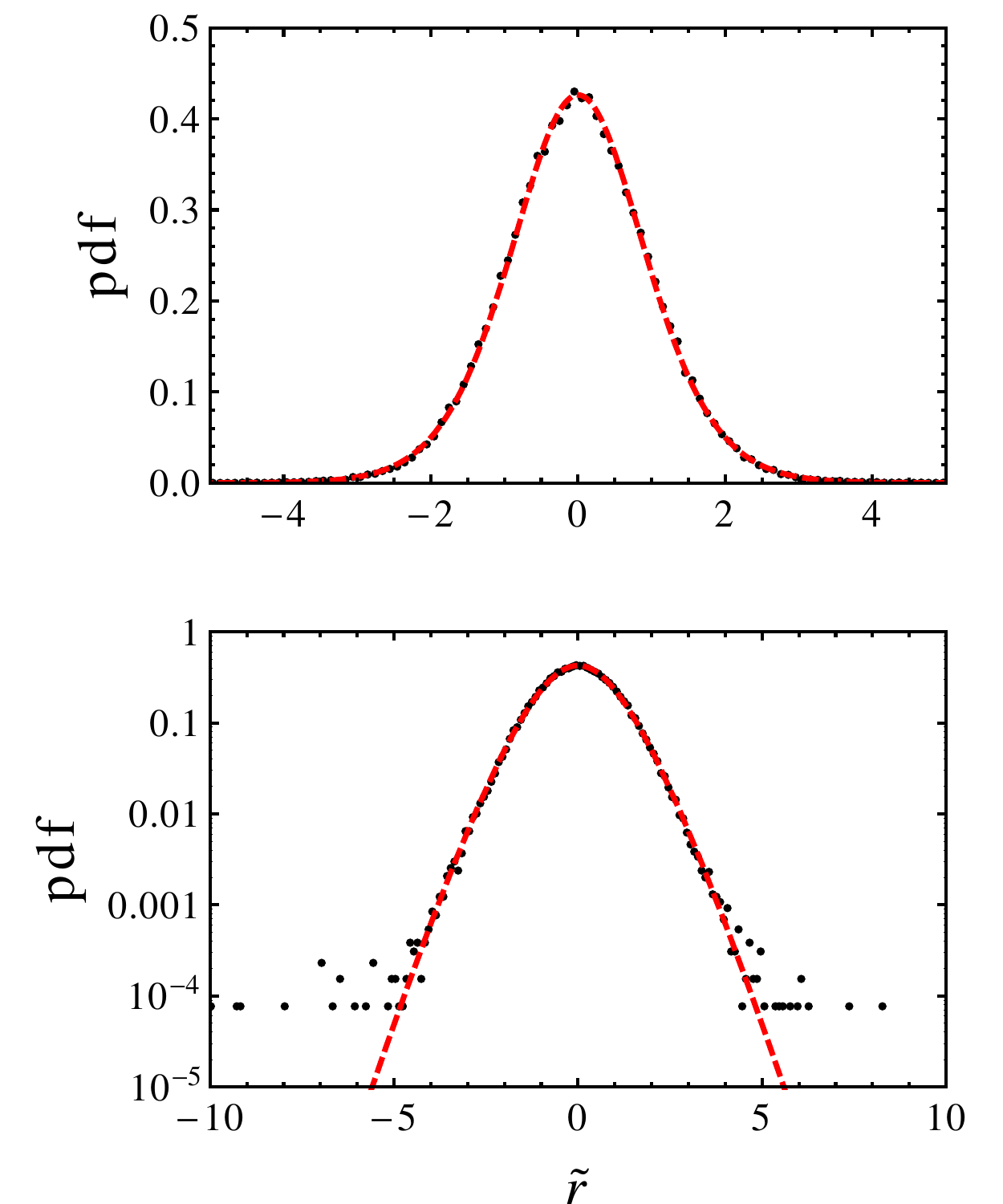}}
\\ 
\centering
\subfloat[$1992-2013$ : $N=4.2$]{\includegraphics[width=\textwidth]{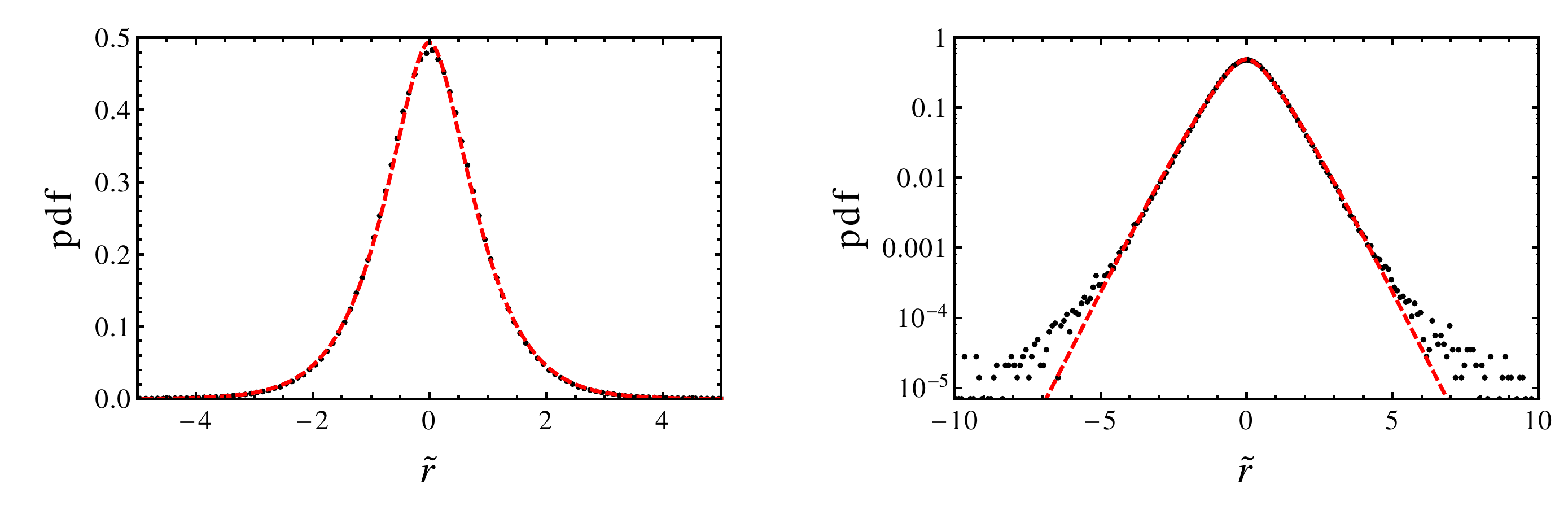}}
\caption{ (a)--(f) Histograms of the rotated and rescaled returns for each market state and (g) for the whole observation period $1992-2013$ compared with the average return distribution (\ref{result}).}
\label{fig6}
\end{figure*}

\begin{figure*}[h]
\begin{minipage}{6cm}
\includegraphics[width=\textwidth]{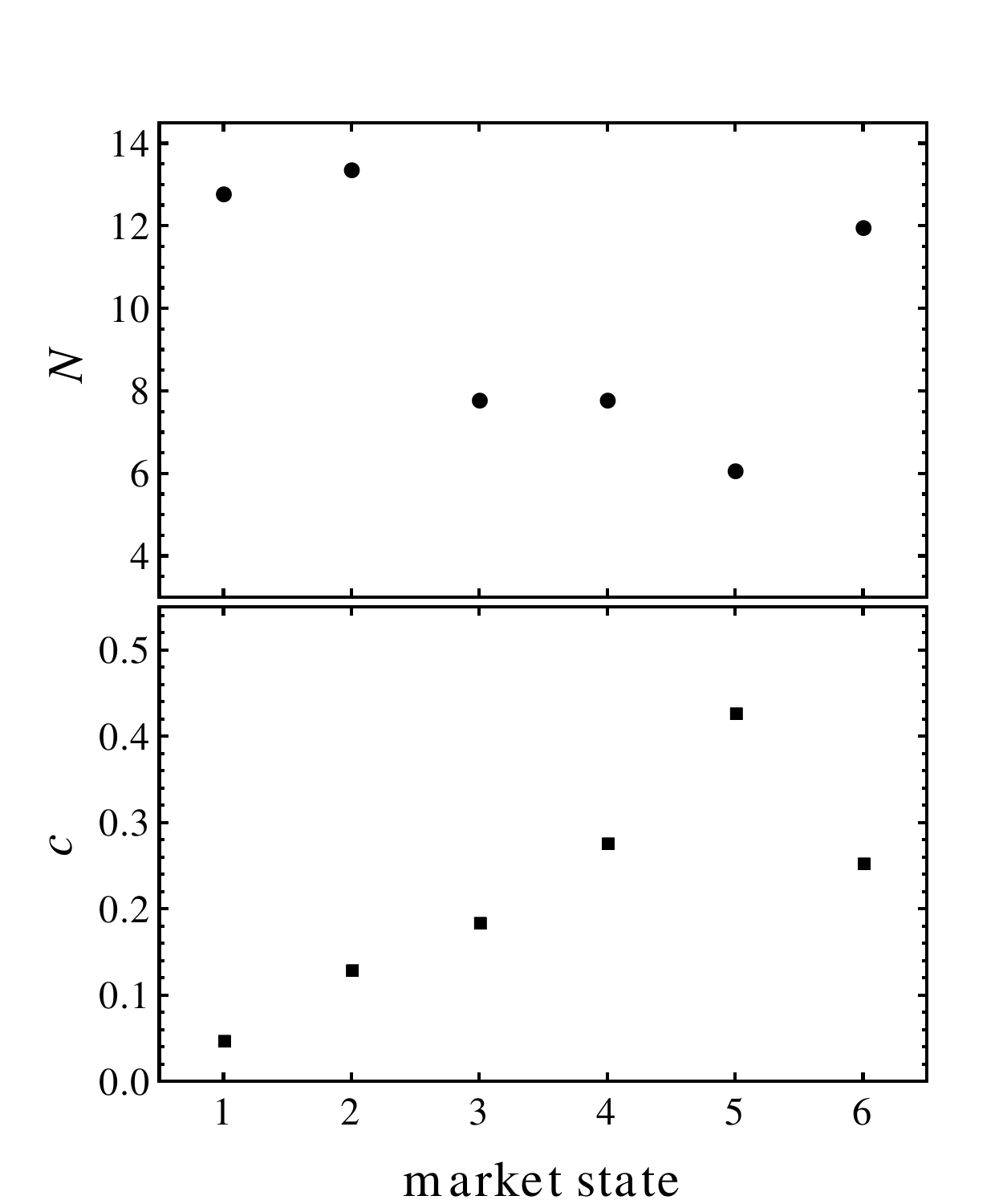}
 \begin{center} \footnotesize (a)  \end{center}
\end{minipage}
\quad
\begin{minipage}{6cm}
\includegraphics[width=\textwidth]{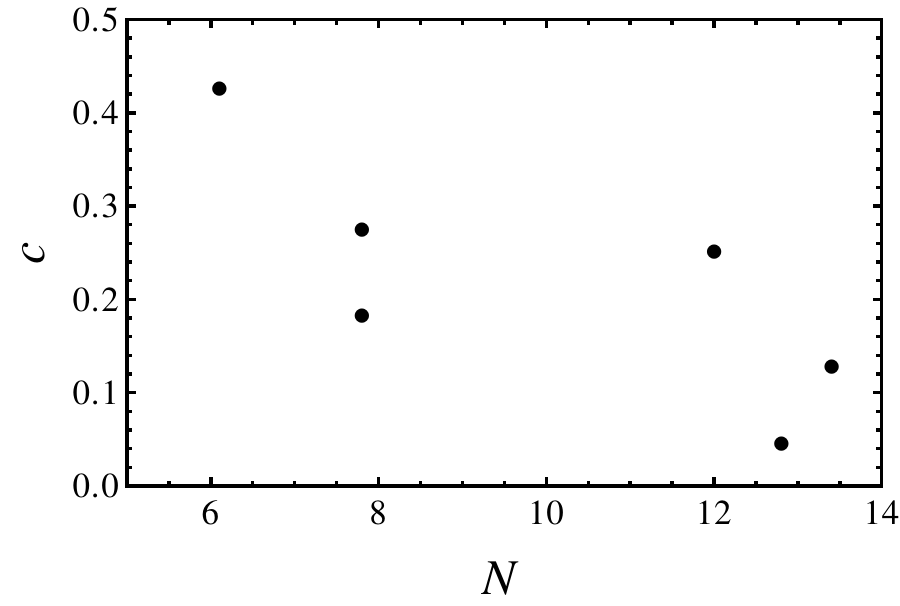}
\begin{center} \footnotesize (b)  \end{center}
\end{minipage}
\caption{ (a) The parameter $N$ (top) and the average correlation $c$ (bottom) for each market state. (b) Scatter plot $c$ \textit{vs.} $N$.} 
\label{fig7}
\end{figure*}

We obtain the return time series for each market state in the following way: 
We take the daily return time series $r(t)$ and divide it into a sequence of disjoint two-month intervals. We merge all intervals belonging to a given state according to the cluster analysis described in section~\ref{section2}. We note that the return time series 
for the 6 market states differ in length. 
For the comparison with the model we rotate the return vector for each state into the eigenbasis of the covariance matrix and normalize its components with the eigenvalues. We aggregate all components into a single histogram and compare it
with the average return distribution~(\ref{result}). Figure~\ref{fig6} shows the results for each market state (a-f) and for the whole $22$-year observation period (g). 
It is already obvious from the heavy-tailed empirical return distributions that the correlations of a given state are not stationary but fluctuate 
around the average correlation matrix.
For the whole observation period we find a much smaller $N$. In this case, the fluctuations are stronger than for the single states. 
The parameter $N$ is estimated by the maximum likelihood method and depicted for each state together with the average correlation $c$ in figure~\ref{fig7}(a). We obtain $c$ by averaging over the off-diagonal correlation coefficients $C_{kl}, \ k \neq l$ of the 
average correlation matrix for a given state.
We observe that the states 1 and 2, which cover the period 1992 to roughly 2002, are rather stable. We find low average correlation with high $N$ values, i.e., weak fluctuations.
In state 3 and 4 the fluctuations increase. While the $N$ values for both states are equal, the average correlation is rising. In state 5, first appearing during the crisis in 2008, the fluctuations increase further. It is the most unstable state with the smallest $N$ value and the highest average correlation $c$. 
In state 6 the fluctuations and the average correlation decrease, the market stabilizes.
To examine the relationship between average correlation and fluctuations we look at the scatter plot between $c$ and $N$, see figure~\ref{fig7}(b). We observe a clear decreasing trend, i.e., a negative correlation. 

\begin{figure*}[h]
\centering
\includegraphics[width=0.6\textwidth]{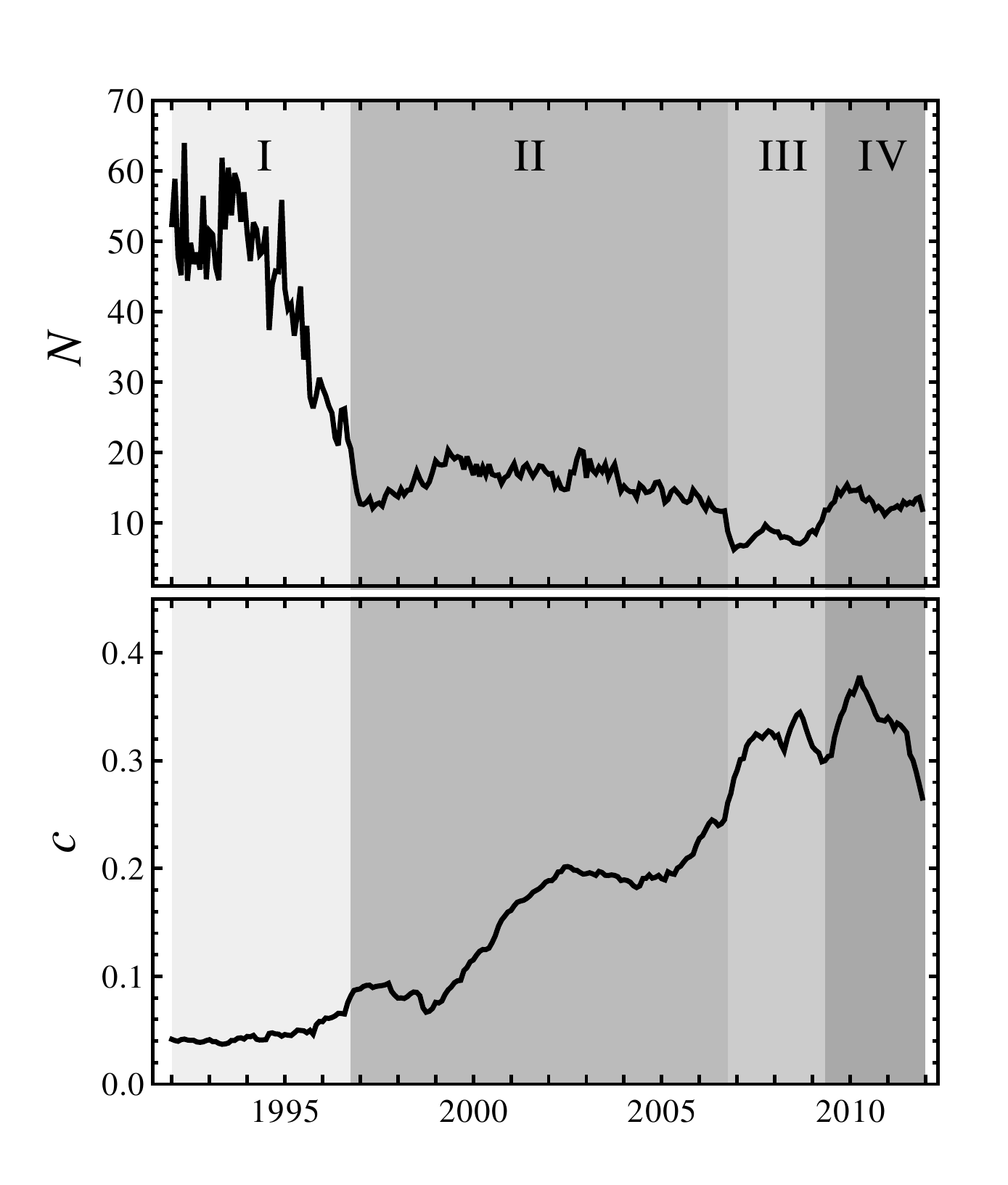}
\caption{Time evolution of the parameter $N$ and the average correlation $c$ calculated on a sliding window of 500 trading days shifted by 21 trading days. The plots are divided into four regimes indicated by different gray scales.}
\label{fig8}
\end{figure*}

To further investigate this relationship we now take a closer look at the variation of the correlation structure over time. To this end, we examine the  parameter $N$ and the average correlation $c$ computed on a sliding window of 500 trading days shifted by 21 trading days, see figure~\ref{fig8}.
We recognize four distinct regimes: The first regime covers the period 1/1992 to 9/1996, which mainly corresponds to the stable market state 1, cf. figure~\ref{fig1}. Here we find the lowest average correlation and the highest $N$ value. 
While the average correlation is relatively stable in this period, the $N$ value shows a clear decreasing trend indicating increasing fluctuations. The second regime covers the period 10/1996 to 9/2006, which corresponds to the stable states 1 and 2 and the more unstable states 
3 and 4. 
While the average correlation in this period is steadily growing, the $N$ value is mostly stable. Compared to the first regime we find smaller $N$ values because of the transitions between the different market states. The third regime, beginning 10/2006 to 5/2009, 
covers mostly the period before and during the financial crisis in 2008 and corresponds to the unstable states 4 and 5. We observe a sharp increase in the average correlation, which is over two times larger compared with the first regime.
Indeed, in times of market instabilities collective behavior is induced which results in larger correlations.  Here we find the smallest $N$ values, i.e., the strongest fluctuations. The last regime, beginning 6/2009, covers the rest of the observation period and corresponds to the unstable 
states 4 and 5 and the stable state 6.
The fluctuations decrease slightly. The average correlation increases at first even further compared to the previous regime but decreases again after 2010. The market stabilizes after the crisis. 

\begin{figure*}[h]
\centering
\subfloat[]{\includegraphics[width=0.5\textwidth]{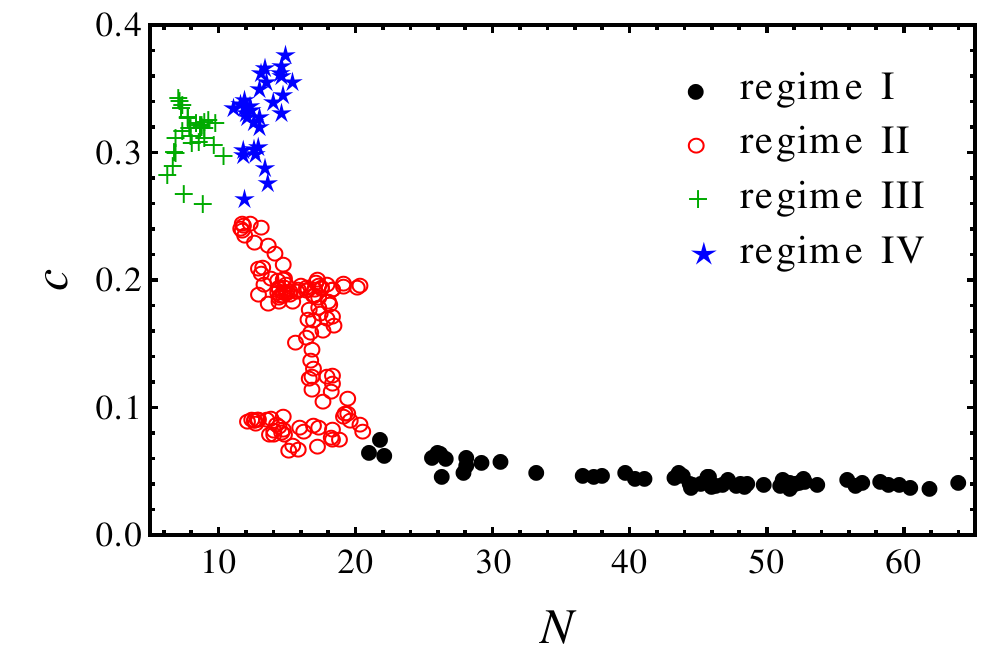}}
\subfloat[]{\includegraphics[width=0.5\textwidth]{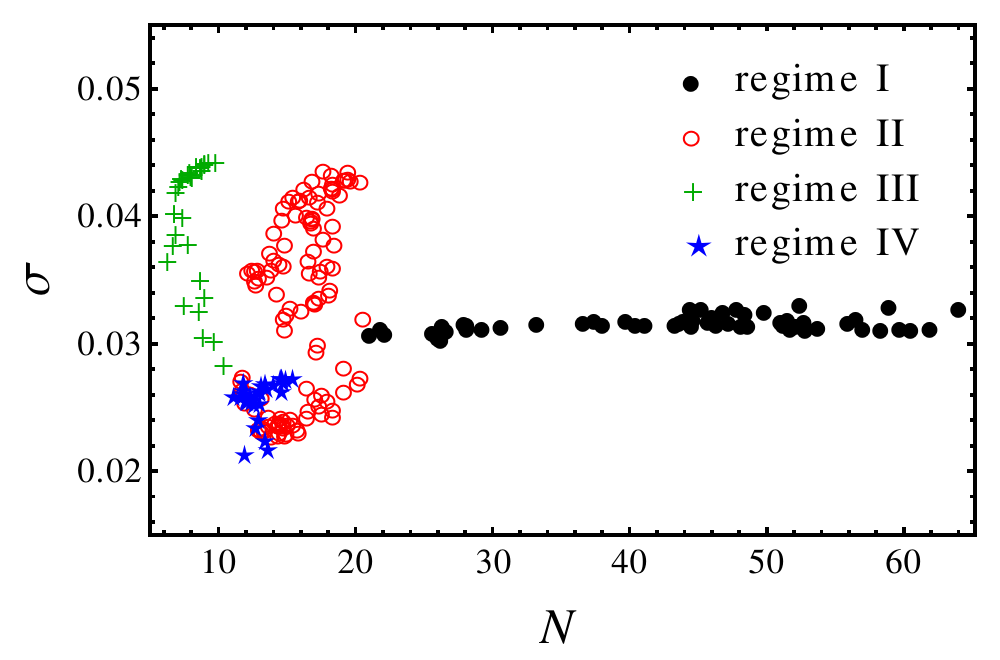}}
\caption{ Scatter plots (a) $c$ \textit{vs.} $N$ and (b) $\sigma$ \textit{vs.} $N$, for the time window of 500 trading days.}
\label{fig9}
\end{figure*}

The relationship between average correlation and fluctuations for the two-year time window is depicted in figure~\ref{fig9}(a). We observe an overall 
negative correlation between $c$ and $N$. 
Moreover, the data corresponding to the four regimes cluster into different regions: a stable region (regime I) characterized by low average correlation and weak fluctuations, which are typical for calmer periods; an unstable 
region (regime III) characterized by high average correlation and strong fluctuations, typical for crisis periods; and an intermediate region (regime II and IV) characterized by varying average correlation and more moderate fluctuations. 

Finally, we examine the dependence between average volatility $\sigma$ and fluctuations for the two-year time window, shown in figure~\ref{fig9}(b). Again, we find clustering into regions as observed before: a stable region with weak fluctuations and nearly constant volatility 
$\sigma \approx 0.03$; an unstable region with strong fluctuations and high volatility; and an intermediate region. In this case, we do not recognize a clear trend, $\sigma$ and $N$ show no clear dependence. 
This justifies our interpretation of $N$ as correlation rather than covariance fluctuations.

\section{Conclusion}  \label{section5}

To achieve a better understanding of the financial market, the concept of market states as clusters of correlation matrices was introduced in \cite{Muennix2012}. Here, we take a closer look at the statistics of 
market states studying the Nasdaq Composite market over a period of 22 years.  For this  purpose,  we use a recently developed random matrix approach, which models the non-stationarity of correlations by a random matrix ensemble. 
Alongside with a heavy-tailed distribution for the stock returns, the approach provides a method to study the correlation structure by estimating the fluctuation strength of correlations in a given time period directly from the empirical return distributions.

Our study provides a better understanding of the market state dynamics as well as of the stability of the corresponding correlation structure.  Despite the non-stationarity of the market we find a set of quite stable
states in which the market operates. We discuss their statistical properties and study the dynamics of the correlation structure in the whole observation period using a sliding window analysis. We find four distinct regimes with different 
statistical behavior.
The analysis reveals a remarkable relationship between average correlation and fluctuations.
Strong fluctuations most likely occur during periods of high average correlation. Unstable periods are thus characterized not only by larger correlations and high volatilities but also by strong correlation fluctuations. 
Furthermore, we study the relationship between fluctuations and average volatility. In this case, we do not find a clear trend, volatility and fluctuations are mostly independent of each other.

Another, conceptual aspect of the present study should be mentioned. At first sight, the following two results might appear contradictory: when studying the entire time interval of 22 years from 1992 to 2013, we identify, on the one hand,
a small number of distinct states in which the market operates, but on the other hand, we claim that the return distribution for this entire time interval can be modeled by a random matrix ansatz \cite{Schmitt2013}. The simultaneous existence of few distinct states and
of an ensemble of homogenously distributed correlation matrices in the random matrix ensemble might seem incompatible. Importantly, these two features can coexist. The present study may be viewed as a refined resolution of the
really existing (random) matrix ensemble in terms of a superposition of sub-ensembles around the distinct states. Certainly, one might come up with statistical observables that can make this fine structure of the ensemble visible -- as we do in the
present study. However, the plain return distribution itself is a highly relevant quantity, see e.g., \cite{Schmitt2014}. To study it, the data are aggregated, i.e., represented in the eigenbasis of the mean covariance matrix for the entire time interval. This involves
a rotation of the return vector with an orthogonal matrix and thus further randomization which is just an additional averaging over the sub-ensembles. This is why our random matrix ensemble works inspite of the existence of distinct market states.

Such effects are quite common in random matrix models and one of the reasons for their remarkable robustness. Wigner's original random matrix ansatz \cite{Wigner1967} based on a rotation invariant and homogenous ensemble was heavily criticized
by many nuclear physicists. They argued that no realistic Hamilton matrix of a nucleus, calculated in some basis, will look like a random matrix, because it will contain many strict zeros due to selection rules. Thus, there was a blatant disagreement with the idea
of a rotation invariant ensemble. This then led to the embedded random matrix ensembles \cite{Brody1981} which correctly incorporate the selection rules. Nevertheless, the statistical observables of particular interest are indistinguishable for Wigner's original ansatz and
for the embedded random matrix ensembles.

\section*{References}

\providecommand{\newblock}{}

\end{document}